\documentclass[journal]{IEEEtran}

\usepackage{graphicx}
\usepackage{url}
\usepackage{hyperref}
\usepackage{amsmath}
\usepackage{amssymb}
\usepackage{bm}
\usepackage{xcolor}
\usepackage{soul}
\usepackage{multirow}
\usepackage{subfigure}
\usepackage{fancyhdr}

\fancypagestyle{firststyle}
{
   \fancyhf{}
   
   \fancyhead[C]{\scriptsize{This is the post-print of the following article: G. Araniti, A. Iera, A. Molinaro, S. Pizzi and F. Rinaldi, ``Opportunistic Federation of CubeSat Constellations: a Game-Changing Paradigm Enabling Enhanced IoT Services in the Sky," in IEEE Internet of Things Journal, doi: 10.1109/JIOT.2021.3115160. \\Article has been published in final form at: \url{https://ieeexplore.ieee.org/document/9547348}. \\Copyright~\copyright~2021 IEEE.}}
}

\ifCLASSINFOpdf
\else
\fi
\begin{document}
%
\title{Opportunistic Federation of CubeSat Constellations: a Game-Changing Paradigm Enabling Enhanced IoT Services in the Sky}
%
%
%

\author{G.~Araniti, A.~Iera, A.~Molinaro, S.~Pizzi, and F.~Rinaldi 
\thanks{G. Araniti, A. Molinaro, S. Pizzi, and F. Rinaldi are with the Department DIIES, University Mediterranea, Reggio Calabria, Italy and with CNIT - National Inter-University Consortium for Telecommunications, Italy.}
\thanks{G. Araniti is also with Peoples’ Friendship University of Russia (RUDN University).}
\thanks{A. Iera is with the Department DIMES, University of Calabria, Arcavacata di Rende (CS), Italy and with CNIT - National Inter-University Consortium for Telecommunications.}
}

\maketitle

\begin{abstract}
Internet of Space Things (IoST) is a challenging paradigm, which is currently attracting great interest from the scientific and industrial communities. IoST is based on the integration of the space segment into the global Internet of Things (IoT) infrastructure. In the relevant literature, reference is generally made to multiple constellations of nanosatellite platforms, used to enable IoT services on a global scale, including also disadvantaged and poorly infrastructured areas. In this paper, we focus on multi-tenant IoT scenarios, wherein multiple CubeSats constellations are enabled to offer services by exploiting a dynamic federation model. The objective is to efficiently provide services in an IoST scenario by leveraging an effective cooperation strategy originally designed for terrestrial IoT networks, the Mobile-IoT-Federation-as-a-Service (MIFaaS) paradigm. We extend this vision to IoT satellite networks in order to allow a constellation of satellites to effectively execute tasks through a tight cooperative behaviour with other CubeSats constellations. The reported performance evaluation studies show that better performance, in terms of percentage of tasks successfully completed, can be achieved through the implementation of the proposed cooperation paradigm. 
\end{abstract}


\thispagestyle{firststyle}

%
\IEEEpeerreviewmaketitle

\section{Introduction}

The new frontier of the Internet of Things (IoT) is in space. There is no doubt that as in the past the non-terrestrial segment was used to achieve full worldwide connectivity, similarly, in the future we will not be able to do without space technology if we intend to offer IoT services on a global scale. 
The reasons why the Internet of Things has the potential to represent a revival of attention to satellite technologies are manifold. Among others, as highlighted in \cite{cianca_15}, if one needs to offer typical IoT services in disadvantaged and poorly infrastructured areas, or if one needs to connect a multitude of things distributed over large areas to the global network, the use of satellite is the most advantageous solution (i.e., the Internet of Space \cite{chong}). It is therefore not by chance that the integration of terrestrial and non-terrestrial networks is the most cost-effective solution whenever the deployment of next-generation terrestrial coverage in support of massive Machine Type Communications (MTC) is prevented in particular areas of the globe by physical constraints \cite{Leo-Sat}.
In this perspective, both the standardization and research activities are focusing on finding solutions to support non-terrestrial networks in fifth generation (5G) New Radio systems \cite{3gpp1}.

For their part, numerous satellite vendors and satellite service providers are making or have announced that they will make significant investments in the IoT Satellite sector. Swarm Technologies, Hiber, Echostar Eutelsat, are just a small example of companies that have announced their willingness to launch satellites with the specific purpose of supporting IoT services and IoT networking by offering two-way connections to IoT devices distributed worldwide. The offer of IoT and Machine-to-Machine (M2M) connections even in remote areas of the world via a Low Earth Orbit (LEO) satellite network is provided by ORBCOMM, and a similar purpose is declared by the satellite operator Thuraya, while Iridium offers, through its CloudConnect platform, a tool for developers seeking a singular communications platform, leveraging the Iridium network, to manage connected devices worldwide.

In this context, one of the most interesting novelties that has recently appeared on the scene is undoubtedly represented by the family of miniaturized satellites, called CubeSats because of their basic 10x10x10 cm\textsuperscript{3} cubic shape \cite{Woellert}. Initially conceived in universities for teaching and research purposes in different fields, the idea of using CubeSat platforms for global connectivity, focusing on some of the fundamental issues related to communications engineering, has been introduced and discussed in the literature for some years \cite{Almon} \cite{puig}.
However, only recently the main interest has shifted towards studies aimed at making CubeSats, or in general nanosatellites, an integral part of the new generation of integrated terrestrial-satellite network platforms characterized by the use of virtualization paradigms, such as Software-Defined Networking (SDN) and Network Function Virtualization (NFV), mainly for network resource utilization improvement and operating cost reduction \cite{sdn-nfv}. 

In this paper, we intend to contribute to this research line by referring to a satellite platform specifically designed for being integrated into 5G IoT systems and, therefore, empowered with  virtualization technologies of both network functions and device resources, and enabling cloud computing capabilities at the edge of the network \cite{tarik18}.
In a 5G satellite-IoT platform of this type, specifically designed to also include CubeSats, we will focus on a multi-tenant scenario in which multiple CubeSat constellations are provided with the means to offer services in a highly efficient and scalable fashion. 

In this platform, we will also introduce and analyze the performance of a paradigm explicitly designed to enable cooperation of multiple CubeSat constellations, also owned by different providers, and their dynamic federation in order to increase flexibility and effectiveness in providing IoT services.
This is obtained by leveraging an enhanced cooperative paradigm among satellite-IoT objects, according to which opportunistic federations of Virtualized CubeSat (ViCubeSat) Constellations are orchestrated and managed by terrestrial Edge nodes. The term ``opportunistic" implies the purpose of dynamically creating coalitions among CubeSats constellations  whenever this is actually considered advantageous and only if the establishment of each coalition results in an increase in utility for the various players (i.e., the constellations involved). Specifically, in this paper we focus on maximizing the number of executed tasks.

In particular, any ViCubeSat Constellation is considered as a pool of satellite-IoT and terrestrial devices owned by the same provider and virtualized in a Private Cloud. 
This paradigm is derived from the one named MIFaaS (Mobile-IoT Federation-as-a-Service), first introduced in \cite{mifaas1}. 
Compared to classic collaborative strategies in Mobile Cloud Computing (MCC) \cite{4}, the opportunistic federation process we implement does not allow for opportunistic cooperation among single devices, rather it involves ViCubeSat Constellations as a whole and the associated sets of virtualized resources, so to extend the opportunities of reciprocal support in delivering heterogeneous IoT services under typical low-orbit satellite operational conditions.

This paper is organized as follows. In Section \ref{state_of_art}, we introduce the state-of-art, include the main related literature on satellites and virtualization in 5G, and highlight the motivation behind this work. In Section \ref{proposal}, we discuss our solution for federating ViCubeSat constellations and, further, in Section \ref{motivations}, we outline the major benefits introduced by virtualizing and federating CubeSats in 5G IoT systems.
In Section \ref{architecture}, we describe the designed IoT-Satellite Edge architecture consisting of virtualized Cubesats and discuss the federation formation problem. Then, the performance results are analyzed in Section \ref{Performance_evaluation}. Finally, conclusions are drawn in Section \ref{Conclusions}.

\begin{table}[htpb]
\caption{Abbreviations and acronyms.}
\label{tab:note}
\scriptsize
\begin{center}
\begin{tabular}{|l|l|}
\hline
3GPP & Third Generation Partnership Project\\
\hline
5G & Fifth Generation\\
\hline
AAA & Authentication, Accounting and Authorization\\
\hline
API & Application Programming Interface\\
\hline
ARPU & Average Revenue Per Unit\\
\hline
BS & Base Station\\
\hline
CAGR & Compound Annual Growth Rate\\
\hline
CAPEX & Capital Expenses\\
\hline
CoAP & Constrained Application Protocol\\
\hline
COTS & Commercial Off-The-Shelf\\
\hline
DV & Data Volume\\
\hline
ELF & Extremely Low Frequency\\
\hline
FM & Federation Manager\\
\hline
eMBB & enhanced Mobile Broadband\\
\hline
ETSI & European Telecommunications Standards Institute\\
\hline
FedVCC & Federated ViCubeCloud\\
\hline
GEO & Geostationary Earth Orbit\\
\hline
GoS & Grade of Service\\
\hline
GPS & Global Positioning System\\
\hline
GS & Ground Station\\
\hline
GSL & Ground-to-Satellite Link\\
\hline
GSN & Ground Station Network\\
\hline
HAP & High Altitude Platform\\
\hline
HAL & Hardware Abstraction Layer\\
\hline
HTTP & Hypertext Transfer Protocol\\
\hline
ICP & Internet of Things Cloud Provider\\
\hline
IoST & Internet of Space Things\\
\hline
IoT & Internet of Things\\
\hline
LEO & Low Earth Orbit\\
\hline
LoS & Line-of-Sight\\
\hline
LwM2M & Lightweight Machine to Machine\\
\hline
M2M & Machine-to-Machine\\
\hline
MCC & Mobile Cloud Computing\\
\hline
MEC & Multi-access Edge Computing\\
\hline
MIFaaS & Mobile-IoT-Federation-as-a-Service\\
\hline
MTC & Machine Type Communication\\
\hline
NB-IoT & NarrowBand-Internet of Things\\
\hline
NFV & Network Function Virtualization\\
\hline
NTU & Non-Transferable Utility\\
\hline
NTN & Non-Terrestrial Network\\
\hline
OMS & Open Mobile Alliance\\
\hline
OPEX & Operating Expenses\\
\hline
PD & Physical Device\\
\hline
P-POD & Poly-Picosatellite Orbital Deployer\\
\hline
QoS & Quality of Service\\
\hline
RF & Radio Frequency\\
\hline
RFID & Radio Frequency IDentification\\
\hline
SDN & Software Defined Networking\\
\hline
SISP & Satellite IoT Service Provider\\
\hline
SLA & Service Level Agreement\\
\hline
STK & Satellite Tool Kit\\
\hline
UAV & Unmanned Aerial Vehicle\\
\hline
VD & Virtual Device\\
\hline
VEL & Virtualization Enhancement Layer\\
\hline
ViCubeSat & Virtualized CubeSat\\
\hline
VO & Virtual Object\\
\hline
\end{tabular}
\end{center}
\end{table}

\section{Research Background} \label{state_of_art}

\subsection{CubeSats Technology}

Miniaturized satellites, known as CubeSats, are a promising solution to realize a global satellite network at a much lower cost with respect to traditional satellite infrastructures. Originally exploited for academic purposes, CubeSats represent an effective solution to the drawbacks of traditional satellites, such as long development cycles, high costs, increasing congestion, lack of sequential redundancy, and high-risk exposure \cite{Aky1}. 
In fact, their relatively low total cost, together with the short time required from development to deployment, make CubeSats really attractive for actually realizing the global connectivity envisioned by the IoT paradigm.

CubeSats have a modular structure. Starting from the basic cubic module of 10x10x10 $cm^3$ dimension, denoted as 1U, spacecrafts of 2U, 3U, and 6U can be built, with a mass of a few kilograms at maximum.  
They do not contain any propulsion system and are mostly built with Commercial Off-The-Shelf (COTS) components. Thus, the deployment stage is carried out through a Poly-Picosatellite Orbital Deployer (P-POD) system, and they are launched in orbit as secondary payloads.

The available frequency bands for CubeSat-to-Ground communication are the S-band (2-4 GHz), C-band (4-8 GHz), X-band (8-12 GHz), Ku-band (12-18 GHz), and Ka-band (26.5-40 GHz). Regarding the altitude with respect to ground, they are placed in the exosphere at low Earth orbit (altitudes of 500 Km and above). CubeSats are equipped with solar panels and high-capacity batteries, since the sun is their major energy source.

A plethora of payloads can be incorporated inside a CubeSat, ranging from several typologies of cameras to telescopes and sensors for space environment monitoring. As an example, the work in \cite{Aky} defines a new CubeSat for dynamic spectrum satellite communication networks, where payloads consist in cameras and Global Positioning System (GPS) receiver. Moreover, it contains a specific communication system composed of multi-band antenna arrays, photonics-based Radio Frequency (RF) front-end and electronics to support enhanced reconfigurable multi-band radios. Another example is given in \cite{cubesat}, which defines a CubeSat, named QuakeSat, for earthquake signature detection. Its mission is to detect, record, and downlink Extremely Low Frequency (ELF) magnetic signal data to support earthquake activity prediction. The satellite has 650 Km sun-synchronous orbit, and it is equipped with COTS components proposing a low cost small-satellite alternative. 

A holistic overview of various aspects of CubeSat missions and a thorough review of the topic from both academic and industrial perspectives are given in \cite{survey}.

Some recent research studies have set themselves the goal of defining new architectures, based on emerging networking paradigms, which involve CubeSats with relevant roles both in the access network and in a space core-network. Interesting examples are given by \cite{Aky1} and \cite{134}.

Authors of \cite{Aky1} envision a CubeSat network named Internet of Space Things (IoST), which represents a paradigm-shift network architecture to leverage the cyber-physical system with another degree of freedom in the space for global process control and optimization. In IoST, CubeSats play the twofold role of nodes of a network infrastructure providing globally scalable connectivity, and of passive and active sensing devices.
A similar architecture, called SoftSpace, has been proposed in \cite{134}, consisting of four segments, a user segment, a
control-and-management segment, a ground segment, and a
space segment.

To date, CubeSats have been deployed in LEO orbits because this choice reduces costs and implementation complexity.
However, the main problem with CubeSats is the same already addressed in the literature in case of LEO satellites and High Altitude Platforms (HAPs), when these elements are used to complement terrestrial networks \cite{Ara-HAP}.
In fact, in addition to the limitations related to the nature of the channel, there are strong issues due to the orbit nature. LEO satellites operate at lower orbits than Geostationary Earth Orbit (GEO) satellites; thus, their footprint is much smaller and they do not appear to be still with respect to the Earth surface. Consequences are limited transmission-reception time windows, discontinuous availability of the Cubesats with a consequent Grade of Service (GoS) and Quality of Service (QoS) degradation, and a raised complexity in managing the whole system. This is a problem that we also try to face through our proposed federation paradigm, as better illustrated in the remainder of this paper.

\subsection{Satellites in 5G and IoT platforms}

Studies to integrate satellites into terrestrial network infrastructures that have appeared on the telecommunications scene from time to time have always shown that the presence of a satellite segment can enhance network operations under many ways and it clearly represents an added value. Even today, in the 5G era, interest has not diminished; on the contrary, there has been a surge thanks to the enormous attention paid to the pervasive provision of IoT services. As a consequence, standardization bodies are studying how to integrate the so-called Non-Terrestrial Networks (NTNs) into 5G systems \cite{3gpp1}, \cite{3gpp2}. Features of NTNs defined by the Third Generation Partnership Program (3GPP) and their potential in satisfying user expectations in 5G \& beyond networks are reviewed in \cite{survey_ntn}.

By starting from the architectural options currently being discussed in the standardization fora, the authors of \cite{5G_Sat} assess the impact of the satellite channel characteristics on the physical and Medium Access Control layers, both in terms of transmitted waveforms and procedures for enhanced Mobile BroadBand (eMBB) and NarrowBand-Internet of Things (NB-IoT) applications for 5G systems including NTN. Instead, in \cite{zimmer}, the main challenges introduced by the satellite channel in the NB-IoT random access procedure are investigated as well as new solutions and research directions are pointed out to overcome those challenges. In \cite{comp_24}, the effect of NTN inclusion into mobile systems is analyzed through an experimental comparison in term of Key Performance Indicators, such as channel quality index, modulation coding scheme index, downlink throughput, frame utilization, and resource block utilization. 

The satellite air interface evolution, in light of new initiatives considering satellite networks as an integral part of the 5th generation of communication networks, is the subject of several papers in the literature \cite{alagha19}, \cite{SAGIN}, \cite{SAT5G}, \cite {NTN_NR_Ericsson}. Besides, in \cite{vol19}, the main issues related to the integration of the satellite into 5G are addressed with particular attention to a testbed demonstration with the main goal of evaluating how satellite networks can be best integrated within the convergent 5G environment.

Obviously, satellites are also studied in the literature for their role of backhauling in 5G networks; an architecture framework with such a finalization is designed in \cite{wang18}. 
In such a context, it is worth also highlighting the research described in \cite{Frank}, wherein the authors studied and described the realization of a Cloud-RAN in a system based on Unmanned Aerial Vehicles (UAVs), acting as mobile Base Stations (BSs), and CubeSats when the scope is the realization of reliable wireless networks in complex areas.

\subsection{Virtualization in IoT and Satellite-IoT platforms}

Obviously, any investigation concerning the inclusion of new-generation satellites in future communication infrastructures for IoT (or for MTC) can not ignore two design aspects characterizing emerging 5G systems: \textit{device virtualization} and \textit{network programmability} (via SDN/NFV), aimed at effectively orchestrating services \cite{Taleb1} and slicing  the network infrastructure \cite{Foukas}.

Modern device virtualization paradigms have overwhelmingly entered also the IoT world owing to the numerous advantages they promise if used in any platform specifically designed to offer IoT services. These include the capability of meeting the stringent timing requirements of many IoT systems, distributing data processing across the so-called cloud-to-things continuum, reducing IoT devices complexity and energy consumption by offloading tasks to the infrastructure, providing efficient IoT resource orchestration, and even contributing to the resolution of various IoT security related issues. As a consequence, concepts such as Fog Computing \cite{Bon12} \cite{Mour17}, Multi-access Edge Computing (MEC) \cite{tarik18}, and the Cloudlet \cite{Bil18} have recently attracted the attention of the entire IoT research community. Examples of research works focusing on the role of the cited paradigms in supporting IoT are given in \cite{Yu18} \cite{Chi16} \cite{Nit16}.

As for network resources virtualization and infrastructure slicing, the emerging paradigms are NFV and SDN. The former, under standardization mainly by the European Telecommunications Standards Institute (ETSI), is a network architecture paradigm that uses information  technologies to virtualize entire classes of network node functions (e.g., router or middle-box functions) into building blocks that may be chained together to create communication services \cite{ETSI} \cite{noi-bruschi}.
Several studies have clearly shown the benefits associated with the joint use of NFV and SDN in IoT platforms, and have demonstrated that typical IoT challenges can be effectively faced through the use of these paradigms \cite{Omn18} \cite{Ojo16} \cite{li17} \cite{135} \cite{136}.

The purposes of introducing network programmability and  resource virtualization concepts also in an integrated terrestrial-satellite 5G platform are manifold. In \cite{Ahmed} cloud-based infrastructures and SDN are exploited for flexible high-level resource sharing, while higher network resource utilization, network management simplification, and operating cost reduction are the objectives addressed in \cite{Aky1}, \cite{Bertaux}, \cite{Li}, \cite{Ferrus}, and \cite{guo}. Besides, sharing CubeSat communications, sensing, and actuation resources within an effective service orchestration framework is addressed in \cite{noi-globecom}, whereas forming a shared resource pool with non-standardized aircraft heterogeneous resources by leveraging resource virtualization technology is the objective of the satellite cloud architecture in \cite{new1}.

Another aspect, largely addressed in the literature and particularly relevant to our research, is undoubtedly the role of NFV-SDN in carrying out the chaining of virtualized network functions to compose network services that provide different trade-offs and functionalities to diverse ecosystems, which is the basis of the network slice concept \cite{SAGIN}. The concept of network slice appears even in \cite{dong}, where the capacity performance of a three-layer heterogeneous satellite network is investigated. Network slicing is also explored for the satellite ecosystem in \cite{decola} to boost eMBB services by ensuring different grades of QoS in dynamic scenarios. 

Our paper leverages precisely this flexibility to compose services based on tasks that can be performed by satellite-IoT elements (i.e. CubeSats) belonging to different constellations, also owned by different tenants, suitably federated.

\subsection{Contribution of this work}

In this paper we design a 5G satellite-IoT platform, which is empowered with virtualization and edge computing capabilities.
It includes CubeSats in a multi-tenant scenario, wherein multiple CubeSat constellations, properly virtualized, cooperate among each other and dynamically federate under the control of terrestrial edge nodes in order to provide IoT services on a global scale.

\section{MIFaaS: the Reference ViCubeSat Constellation Federation Paradigm}  \label{proposal}

Our solution for federating ViCubeSat constellations is inspired by the MIFaaS paradigm designed for terrestrial IoT platforms 
that we re-think and adapt to the specific needs of an integrated terrestrial-satellite scenario for 5G IoT. 

The MIFaaS paradigm has been first introduced in \cite{mifaas1} contributing to the evolution from a centralized cloud computing to a hybrid distributed paradigm, extending its scope from the core to the extreme edge of the network. Traditional cloud-like resources (i.e., storage, computing, applications) are provided not only in the remote cloud, but also in the network edge infrastructure and in heterogeneous end-devices, belonging to private users/companies/service providers. Different from the typical device-oriented approach of MCC, where the single device can receive opportunistic support from nearby devices and their embedded resources only, in MIFaaS the attention is on federation of pools of devices/objects managed by private/public owners, the so-called IoT Cloud Providers (ICPs) \cite{mifaas1}.
 
A platform implementing MIFaaS leverages the recent advances in virtualization techniques in the fields of networking (i.e., virtualization of network resources and functions, SDN and NFV), cloud/fog computing (i.e., virtualization of computing and storage resources), and IoT (i.e., object virtualization). Virtualization technologies are used to create network-assisted Virtual Devices (VDs), which are meant to describe the capabilities and features of the corresponding Physical Devices (PDs), belonging to an ICP. They are also capable of complementing and offloading applications/services, high-level network functions and cloud-like resources (e.g., control logic modules, analytic tools, firewall, storage, processing) residing in potentially constrained end-user devices. 

ICP resources can be exposed and accessed as a single cloud, regardless of their actual physical location. Moreover, by exploiting network softwarization technologies, the edge infrastructure will be able to dynamically allocate and seamlessly migrate (all or some of the) VDs associated to an ICP moving them closer to the end-user whenever needed (e.g., in case of mobility) for the fruition of a service with the requested Quality of Experience (QoE) level.

\section{Motivations to federating CubeSat Constellations}  \label{motivations}

\subsection{Benefits of virtualization}

\subsubsection{Managing and orchestrating constellations}

Virtualizing a CubeSat in the ad hoc ``softwarized'' terrestrial segment provides the opportunity to improve the efficiency of the typical monitoring and management procedures of CubeSats, regardless of their current position. A very welcome aspect is in fact carrying out CubeSat management operations  (Bootstrap, Device Discovery and Registration, Device Management and Service Enablement, etc.) in continuity and without having to wait for a complete rotation cycle along the satellite orbit. This can be achieved by associating the CubeSat with a Virtual Object (VO) enabled to migrate from ground station (GS) to GS during the CubeSat revolution around its LEO orbit. In a certain sense, this solution enables tracking the respective hardware device in the sky. In this way, an image of the satellite is always available in the GS to which it is currently connected (i.e., at the edge of the network) and that can continuously interact with the satellite. Besides, if typical IoT virtualization techniques are used, this image and its associated resources would also be accessible remotely by using standard protocols for IoT.

In the solution proposed in this paper, we rely on the  Open Mobile Alliance (OMA) Lightweight Machine to Machine (LwM2M) standard for CubeSat virtualization \cite{oma-registry}.  This standard is targeted, in particular, at constrained devices and provides a light and compact secure communication interface along with an efficient data model, which enables both device management and service provisioning for M2M devices \cite{Prado}. Obviously, any other standard for virtualization can be used without affecting the overall functionality of the infrastructure that we propose.

\subsubsection{Continuous sensing}

Equally interesting is the ability to continuously access satellite data. 
At the moment, you can access the data sensed by the CubeSat only during its stay under the control of the GS and then you have to wait for the rotation time along an entire orbit before having the data available again. In a virtualized and multi-tenant scenario, the VO could also be migrated (following the CubeSat) to land stations of other tenants that, subject to commercial agreements, can temporarily host it. The VO would interact with the CubeSat by using the Constrained Application Protocol (CoAP) to update the data and such a ``roaming'' VO would be reachable via IP network from remote by simply using Hypertext Transfer Protocol (HTTP) commands. 

\subsubsection{Augmented data fruition}

A further advantage could derive from the possibility of making an association (also dynamic) of more CubeSats devices into a single ``Virtual CubeSat'' software. This technique, widely used in different environments where sensing or identification operations are required, such as Radio Frequency IDentification (RFID) EPCGlobal, would make it possible to make the most of a multi-tenant scenario composed of different CubeSats constellations or even a scenario where the single tenant owns several constellations. A classic example is to want an updated datum related to an area at a given time when there is no owned CubeSat passing on that area that can provide the data. In this case, by querying the Virtual CubeSats, data could be transparently provided by any of the real CubeSats associated with it (even belonging to a different tenant, subject to business agreement between tenants) that is currently passing over that area.

\subsection{Benefits of federation}

The main advantage of ViCubeSats federation derives from the observation that there are two typologies of constellations belonging to each tenant: a typology that we will define ``heterogeneous'' and another that we will define ``homogeneous''.

The term ``heterogeneous'', in this situation, indicates that the satellites belonging to the same constellation have different natures and are equipped with different sensing payloads. This implies that the ViCubeSat constellation itself is capable of offering a multiplicity of resources by leveraging on which different type of tasks can be solved. We can observe that, in such a situation, in principle, the ViCubeSat constellation does not have a strict need to cooperate with others, since it already owns the needed resources both associated to the CubeSats (sensing resources) and to the terrestrial segment (as an example, specific computational functions or buffering resources) belonging to the same tenant. 

Despite this, the situation which sees the presence of ``homogeneous'' constellations is also very common, meaning that all the satellites belonging to a constellation offer the same resources. We expect that, in this situation, a ViCubeSat constellation belonging to a tenant is more inclined to share its resources with others offering a different type of resource (belonging to a different tenant), since the resolution of composite tasks could likely require resources it does not own. The federation of ViCubeSat constellations comes into play in this second situation.

\subsection{Business Model for the New Paradigm}

The CubeSat constellation federation paradigm will bring a novel business model with clear key advantages to all the involved players. It will definitely contribute to the growth of Satellite IoT Service Providers (SISPs) by increasing the revenue opportunities, owing to new service offered to end-users and business opportunities with other service providers. SIPS can offer their existing programmable network infrastructure in support of novel value-added services with Capital Expenses (CAPEX) and Operating Expenses (OPEX).

By leveraging on NFV and SDN technologies, SISPs can support and manage the federation of private clouds by offering additional and enhanced services to their subscribers by also increasing the Average Revenue Per Unit (ARPU). As an alternative, they can open their network infrastructure to authorized third-parties, such as Cloud Providers that can flexibly complement the capabilities of their remote data centres with resources at the satellite-terrestrial infrastructure edge to better satisfy their customers and attract new ones, such as emerging companies which have a business interest in rolling-out innovative IoT applications and services towards end-users without owning and operating any physical satellite infrastructure.

SISPs (not only big players but also Small and Medium Enterprises, that is SMEs) would benefit from a higher flexibility in designing novel applications and services by exploiting novel Application Programming Interfaces (APIs) and protocols made available to offer in-network support for service orchestration and delivery to the end users. In doing so, they will be able to offer a broader service portfolio, with increased pervasiveness (by overcoming the deployment limitations) but with limited investments.

According to market researches (Allied Research, https://www.alliedmarketresearch.com/small-satellite-market) on the global small satellite industry, the global small satellite market is expected to reach \$15,686.3 millions by 2026, registering a Compound Annual Growth Rate (CAGR) of 20.1\% from 2018 to 2026. In this context, the nanosatellite (CubeSat) market is expected to experience a high rate of growth due to low construction costs, light weight, ease of construction and development, and high ability to perform complex computational tasks. A further driving factor will be the possibility of making investments that can be faced by small and medium-sized enterprises and startups entering the market.

Obviously, the global market for small satellites is driven by the need to reduce costs related to space missions.
Unfortunately, the technologies to access space, despite being consolidated, require a high level of technological specialization, a long experimentation, and are expensive. This is why national space agencies have started collaborations with the private sector to reduce costs and increase the technological level of products, whether it is satellites or launchers \cite{geoforall}.

There are two possible choices to obtain a cost reduction. The first involves choosing between carriers already on the market, capable of handling medium-small payloads but built and managed by the main space agencies. In this case, it is possible to take advantage of both multiple launches of an adequate number of small satellites, or a ``piggy-back'' type launch, where the smaller satellites are launched together with a main load considerably sharing the launch period, the site of departure, vector and type of orbit.

The second involves the use of a smaller carrier but developed for this sector, perhaps starting from sounding rockets.

As mentioned above, in this paper, our proposal aims to give the CubeSats market further impetus thanks to a third alternative, which provides for the possibility of reducing the launches of small satellites and instead exploiting a disruptive business model for the shared use of CubeSat constellations following the establishment of appropriate federations, as it will be better explained below.

\section{The reference architecture} 
\label{architecture}

\subsection{Functional model of the designed IoT-Satellite Edge architecture}

Our platform is based on the concept of hybrid distributed cloud, extending its scope from the core to the extreme edge of the network. To this aim, traditional cloud-like resources (i.e., storage, computing, applications) are provided not only in the remote cloud, but also in the network edge infrastructure (i.e. the GSs, where fog computing is expected to be supported) and in heterogeneous end-devices (i.e. CubeSats with different technological characteristics and different payloads), belonging to different private tenants (Satellite Service Providers) or deployed by public authorities (Space Agencies, Research Centers, Universities, etc.).

The platform enables the federation of different ViCubeSat constellations, ViCubeClouds (i.e., resources virtualized from physical devices belonging to the same owner), so that a tenant offering cloud satellite services may benefit from the availability of resources from physical devices deployed on the ground and from the CubeSats belonging to different constellations. 

The keys to support the federation are the latest IoT object virtualization technologies, which are used to create ViCubeSat meant to describe the capabilities and features of the corresponding CubeSats. For achieving an effective integration with future 5G terrestrial platform, the platform also leverages the recent advances in virtualization techniques in the fields of networking (i.e., virtualization of network resources and functions, SDN and NFV, respectively), cloud/fog computing (i.e., virtualization of computing and storage resources). 

All together, these approaches contribute to consolidate distributed resources of a ViCubeClouds in a virtual manner, so that they can be exposed and accessed regardless of their actual physical location. Moreover, by exploiting network softwarization technologies, the edge infrastructure will be also able to dynamically allocate and seamlessly migrate (all or some of the) ViCubeSat resources associated to a ViCubeCloud, moving them closer to the place where they are needed for the fruition of a service with the given QoE level.

We have to consider that there is a fundamental difference between the terrestrial scenario and the satellite scenario: the CubeSats follow a precise trajectory that respects mathematical laws. This feature allows us to predict the movement of satellites and to a priori know which satellites are in Line-of-Sight (LoS) condition with respect to every GS at each time instant. The federation time is related to the CubeSat-to-Ground link activity time due to the fact that the task must usually be performed during the visibility window. 

Due to the intrinsic nature of a constellation, the edge infrastructure owned by the Tenant is the best candidate to host ViCubeSats and any other type of virtualized resources. In fact, beyond its geographical nature, the edge infrastructure is in charge of terminating secure access connections towards CubeSats, and to actuate Authentication, Accounting, Authorization (AAA) operations across the access network technology.

Our platform foresees the presence of two main software entities in charge of tracking the status of CubeSats and ViCubeSats. In particular, a ViCubeSat Controller is associated to the physical device to monitor the status of the offered physical resources and capabilities. Off-the-shelf utilities provided by end-devices will be used to this purpose, and properly augmented to interact with the corresponding ViCubeSat at the edge infrastructure. A ViCubeCloud controller, hosted at the edge infrastructure is, instead, responsible for monitoring the status of ViCubeSats in the ViCubeClouds. It represents the main interface to enable the federation among different ViCubeClouds of things and services.

Federation creation and management policies are implemented by a broker entity, acting in various points of presence of the terrestrial segment edge infrastructure, according to federation requests received by the ViCubeCloud controllers. Whenever two or more ViCubeClouds are trading and sharing resources, a Federated ViCubeCloud (FedVCC) is created. Decisions on resource sharing policies within a FedVCC and the dynamic federation maintenance are responsibility of the broker, which has a complete knowledge (in terms of status and capabilities) of the available resources in the fog, in the ground and in the sky that can be shared in the federation.

\begin{figure*}[h]
\centering
\includegraphics[scale=0.4]{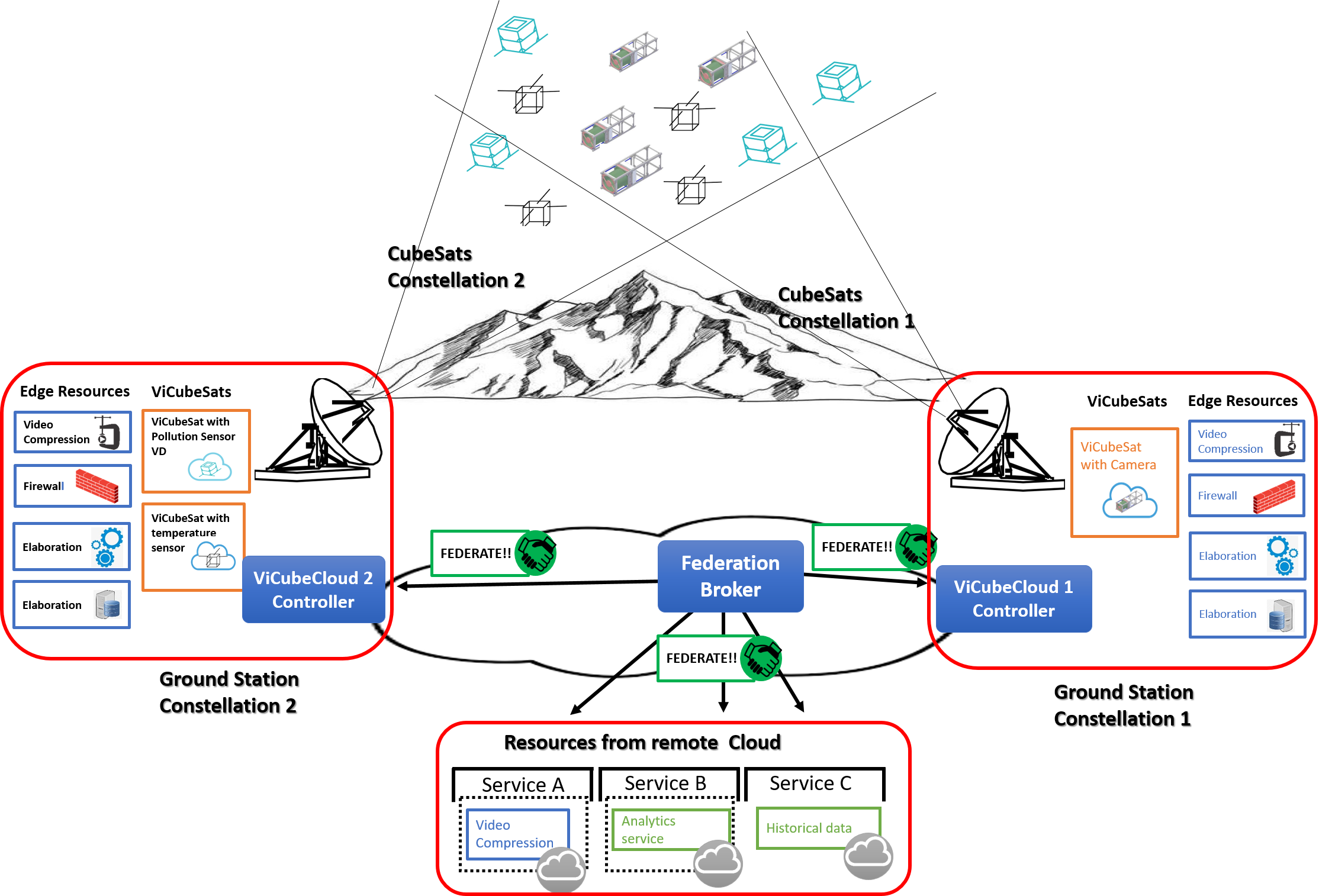}
\caption{Reference architecture.}
\label{fig:architecture}
\end{figure*}

The decisions the broker takes are supported by the Federation Manager (FM) block, implementing policies and defining criteria for an effective and efficient formation and maintenance of federations among distributed heterogeneous private clouds, also according to negotiated Service Level Agreements (SLAs). The conceptual approach of our infrastructure is shown in Figure \ref{fig:architecture}, where a user has to run an application requesting a resource currently unavailable in his/her own ViCubeCloud (i.e., a pollution sensor) and retrieved, instead, thanks to a federation, securely set-up by the federation broker, with another ViCubeCloud.
To achieve its ambitious objectives, the design of the platform does not start from scratch; on the contrary, it will exploit the concepts and achievements developed in the INPUT  project (https://www.input-project.eu/). In particular, the platform will use the features of OpenVolcano \cite{openvolcano}, defined in the INPUT project, which proposes to empower the edge infrastructure with ``in-network" softwarization capabilities to allow networking technologies to be more integrated with cloud services. 

The platform is split into a \textit{control plane} and a \textit{data plane}. The first is composed of:
\begin{enumerate}
    \item The \textit{Data Collection and Configuration Layer}, which provides interfaces to configure and optimize the hardware resources according to the quality levels required by the different services, collects services status, and manages services. 
    \item The \textit{Elaboration Layer}, which \textit{(i)} aggregates and analyzes monitored data, \textit{(ii)} manages monitoring information and subscription of services, and \textit{(iii)} real-time configures logical resources. 
    \item The \textit{Commit Layer}, which actuates the Openflow rules computed in the Elaboration layer and abstracts the virtualization libraries of data plane to make them platform-independent for the elaboration layer.
\end{enumerate}

The data plane is composed by OpenFlow switches and computing facilities that can host both virtual machines and containers (i.e. Docker).

Besides the provision of tools for federating virtual resources belonging to different Virtualized Constellations, some interesting features are enabled by our platform, which go well behind what is available today in the current satellite realm.
Currently, each owner manages its constellation of CubeSats which are usually connected to one GS only. As a consequence, CubeSat-to-Ground communication is limited to the interval during which CubeSat-to-GS link is active. 
Our platform has the potential to enable the creation of a multi-tenant GS Network (GSN) where GSs are edge nodes of the same global terrestrial satellite network. Each tenant is motivated to share its computation and radio resources (following suitable business agreements), host third-party microservices (VOs), and communicate with CubeSats owned also by other tenants when in visibility to the aim of improving the service level offered to their subscribers through ViCubeSat federations.

\subsection{Virtualization of CubeSats}

Within the IoT, virtualizing PDs \cite{Nit16} is a consolidated practice that provides several advantages to constrained objects. Especially in Fog environments wherein PDs are in proximity with their virtual counterparts, VOs can benefit from increased computational resources with respect to the constrained PD hardware.

From the perspective of this work, the PD is a CubeSat with all its equipment of sensors, actuators, and functional components.

The VO contributes to application-independent virtualization of CubeSat. Its services are exposed by HTTP APIs enabling interconnection with several consumers in a secure way.

The CubeSat VO presents two logical API levels: \textit{(i)} the \textit{PD Interface}, which is dedicated to communications with PDs, and \textit{(ii)} the \textit{Application APIs}, which expose a set of enhanced customized APIs to applications that require services from the VO. The VO core, the Virtualization Enhancement Layer (VEL), and the OMA LwM2M Hardware Abstraction Layer (OMA-HAL) allow enhanced virtualization, request management, semantic description, and context awareness.
Our VEL is implemented with improved OMA-LwM2M features and optimized for CubeSats.

The OMA-LwM2M standard defines a client-server protocol where each device is defined by a resource and data model that can be customized to any device. The LwM2M uses standard REST-interfaces provided by the CoAP protocol, providing a request/response interaction model between application end-points and supporting built-in discovery of services and resources \cite{OSCORE}. 
The basic information that a LwM2M Client transmits is a \textit{Resource} data, while \textit{Objects} are composed of a set of resources. Basically, an object is used to describe and control a specific software/hardware component (such as sensors, antennas, or device firmware) with associated Resources (e.g., value, unit, max value, min value). Depending on the object characteristics, we could have more instances of the same object in the device. More details on a LwM2M Client, flexible enough to virtualize CubeSats that can incorporate a rich range of payloads, are available in our previous paper \cite{noi-globecom}.

Both VEL and OMA-HAL rely on the VO datastore, which is hosted in the container Data Volume (DV). Details of each functional module of our VO can be found in \cite{noi-bruschi} where a similar approach has been used for terrestrial-only systems.

It is worth noting that the VO is designed to be PD-independent. This means that the same VO service template can be used to instantiate virtual counterparts of different CubeSats or PDs.

\subsection{Federation policy}

To implement an edge-assisted collaboration among ViCubeSat constellations, we use a game theoretic model for the federation formation problem, which allows to better capture the rationality of the users (i.e., the owners of each constellation) and their willingness to join or leave a federation according to their personal preferences over the utilities they reach. 
The purpose of this paper is not to optimize the choice of constellation coalition policy and we leave this as a starting point for future research. Therefore, in this paper, we use the approach already utilized successfully in the paper \cite{mifaas1} and we obviously apply it to our scenario with the necessary adaptations.
In our scenario, without losing in generality, we assume a GS (the Edge-node of our Edge-ViCubeSat architecture), which can communicate with CubeSats equipped with heterogeneous payloads and belonging to different constellations owned by different Tenants/Providers.

We assume that its main objective is to create and orchestrate one or more ViCubeSat federations so to maximize the number of executed tasks while meeting the constraints set by the different owners of the constellations (self-interested players).

Analogously to \cite{mifaas1}, we consider a network with a set $N$ of ViCubeSat constellations (in the following named ``virtual constellation”), where the $i$-th virtual constellation is endowed with a set of $D_i$ devices. Resources can belong to a set of $M$ different resource types. 
The $i$-th virtual constellation has a workload of task requests assigned by the edge node on the basis of the received application requests.

A further important feature characterizing a virtual constellation is the mobility of its CubeSats and, therefore, the availability of their associated resources in the area of visibility of a given GS. Obviously, all CubeSats belonging to the same virtual constellation have the same mobility pattern. The mobility parameter is of high concern during task allocation, as it influences the probability that a task assigned to a given CubeSat of a federated constellation is performed and the result is timely received. 

We assume an ``edge node assisted'' solution, whereby once a task is allocated to a virtual constellation (i.e., to some CubeSats belonging to this constellation), the data exchange relevant to that task goes over the edge node (the GS in visibility). Therefore, it is important that the CubeSats of federated constellations involved in the task assignment remain under the coverage of the GS and maintain the connection to the corresponding ViCubeSat Controller in the GS for the time required to execute the task and transfer the results. Obviously, being the mobility pattern of the CubeSats around their orbit precisely predictable, the Controller at the GS can exactly know whether or not a CubeSats leaves the area of interest and, hence, loses the contact with its local Controller (i.e. the CubeSat HW moves to the visibility area of the next GS and its Virtual Image migrates to the next GS) before being able to complete the assigned task. This is a feature that simplifies the model used in \cite{mifaas1}, wherein some probabilities that a whole Private IoT Cloud leaves the Edge Node connection were considered.

To model the cooperation problem among the virtual constellations, we use the Non-Transferable Utility (NTU) coalitional game based on the coalitional game theory \cite{gametheory} defined in \cite{mifaas1}. 
For completeness, we briefly recall here that in \cite{mifaas1} we considered an NTU game $(\mathcal{N}; \mathcal{V})$ where $\mathcal{N}$ is a set of $N$ players and $\mathcal{V}$ is a function, such that, for every coalition $\mathcal{S} \subseteq \mathcal{N}$, $\mathcal{V}(\mathcal{S})$ is a closed convex subset of $\mathbb{R}^{|\mathcal{S}|}$. The latter contains the payoff vectors that the players in $\mathcal{S}$ can achieve, and $|\mathcal{S}|$ is the number of members in the coalition $\mathcal{S}$. The objective for the players in the NTU game in \cite{mifaas1} was to maximize their associated value in the coalition they belong to. Such a value is considered as the difference between the gain that the players obtain (i.e., the utility) in cooperation minus a cost term associated to the sharing of their resources. The utility term for any player was associated to the number of tasks being successfully executed in a given area over the number of tasks requested in that area, whereas the cost term was a measure of the amount of resources used by a Private  IoT Cloud in a given coalition to execute all assigned tasks over the total amount of available resources. The interested reader can refer to \cite{mifaas1} for more details of the mathematical expressions of utility and cost.
The same NTU game has been used in this paper, properly adapted to the case in which the players are the ViCubeSat constellations with their virtualized CubeSats.

Obviously, we also need to define a \textit{task allocation algorithm} that maps the set of task requests in the federation with the available resources. Being task allocation not the main focus of this paper, we consider a simple algorithmic solution, based on a greedy approach that assigns tasks by considering the number and the type of resources available in each federation.
More sophisticated and effective algorithms can be easily taken from the literature and used without affecting the validity and the generality of the proposed cooperative paradigm.
In particular, the chosen task allocation algorithm aims at mapping as many task requests as possible to the resources of the involved virtual constellations, compatibly with the resources available at each constellation.

For each coalition, tasks are scheduled by the federation broker, which starts searching for a suitable constellation in the coalition able to provide the needed resources. It first checks the availability of CubeSats in each constellation according to their mobility pattern, which affects the availability time of the CubeSat-to-Ground link (in our case this time interval is exactly known as it is determined by the orbits used by the constellation). Then, in view of achieving higher probability of task execution success and load balancing, it selects the CubeSat that owns the required resource and that has a permanence time within the area of interest compatible with the time required for solving the task. Finally, the ViCubeCloud Controllers update the resource availability status.

\section{Performance evaluation}\label{Performance_evaluation}

In this section, we present a preliminary analysis aimed at evaluating the benefits of our proposed ViCubeSat virtualization platform and federation policy. 

\subsection{Evaluating the virtualization platform}

In this subsection, we discuss the main simulation results regarding the performance of the virtualization platform. In \cite{noi-globecom}, we carried out a more extensive simulation campaign of which we report only partial results. 

\begin{figure}[h]
\centering
\includegraphics[scale=0.3]{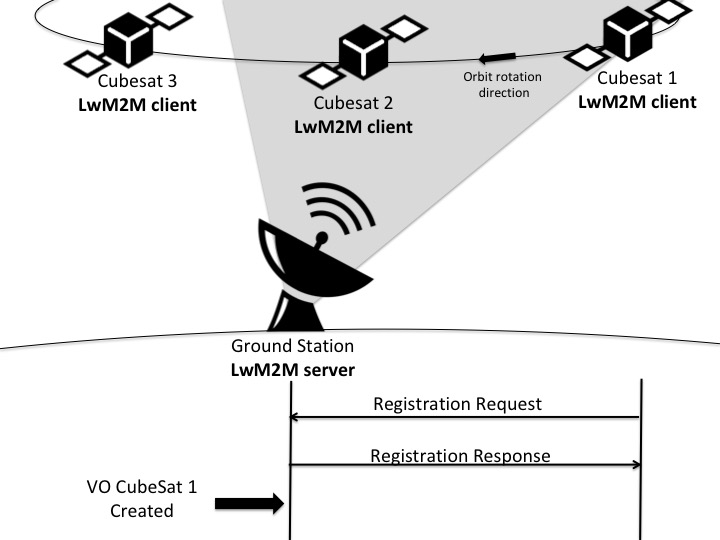}
\caption{Illustration of the OMA-LwM2M Registration procedure.}
\label{fig:VORegistration}
\end{figure}

\begin{table*}[h!]
\caption{Performance of the Virtualization Platform.}
\label{tab:res_virtualization}
\centering{
\begin{tabular}{|c|c|c|c|}
\hline \hline
\textbf{CubeSat Altitude} & \textbf{VO Registration} & \textbf{Potentially deliverable} & \textbf{Potentially deliverable} \\
\textbf{[Km]} & \textbf{Load [\%]} & \textbf{data in DL [Mbytes]} & \textbf{data in UL [Kbytes]} \\ \hline \hline
500	& 0.074 & 85.9748 & 206.3396 \\ \hline
600	& 0.068 & 92.8618 & 222.8685 \\ \hline
700	& 0.065 & 97.5041 & 234.0100 \\ \hline
800	& 0.057 & 106.6663 & 255.9992 \\ \hline
900	& 0.056 & 114.1016 & 273.8439 \\ \hline
1000	& 0.048 & 132.2279 & 317.3470 \\ \hline \hline
\end{tabular}
}
\end{table*}

In particular, we focus on the \textit{OMA-LwM2M Registration} procedure that takes place when a CubeSat enters in visibility of a GS (i.e., the Ground-to-Satellite Link, GSL, is active). This is the case of CubeSat 1 in Fig. \ref{fig:VORegistration}. In such a situation, an handshake between the CubeSat and the GS occurs through the exchange of a \textit{Registration Request} message from the CubeSat that is acknowledged by a \textit{Registration Response} message from the GS. 
After the registration procedure, the server (GS) and the client (CubeSat) are free to exchange data for the duration of the window.

We use the Satellite Tool Kit (STK) \cite{stk}, a software tool that allows simulating Earth-orbiting satellites. Among its feature, STK generates a report of the timestamp of visibilities of the ground to its associated spacecrafts. We exploit this capability to evaluate the GSLs activity time (i.e., access time) when varying orbit altitudes.

According to \cite{LB}, we set the downlink (Ground-to-CubeSat transmission @ 437 MHz) and uplink (CubeSat-to-Ground transmission @ 146 MHz) data rates to 1 Mbps and 2.4 Kbps, respectively. 

In particular, we focus on evaluating the performance metrics in terms of \textit{(i)} \textit{VO registration load}, representing the percentage of time spent, over the access time, for the VO registration, accounting for the transmission of both the \textit{Registration Request} packet from the CubeSat and the \textit{Registration Response} message from the GS, and \textit{(ii)} \textit{Potentially deliverable data}, indicating the potential amount of data that can be delivered in the residual access time after the registration procedure.

Table \ref{tab:res_virtualization} summarizes the simulations results, which highlight that, despite the conservative assumption on uplink and downlink data rate, the VO registration procedure is not a time-consuming task. Indeed, the negligible load of registration allows exploiting almost all access time to deliver data, with a consequent maximum amount of deliverable data equal to about 130 Mbytes and 320 Kbytes in downlink and uplink, respectively.

\subsection{Evaluating the federation policy}

In this subsection, we evaluate the performance that can be obtained when implementing a federation policy in the proposed virtualization platform. The analysis is conducted by using a simulation code developed in MATLAB, which allows to evaluate the impact that the number of virtual constellations, the number of satellites per constellation, the orbital height, the number of required tasks, and the tasks execution time have on the overall achievable performance. 
We consider a generic IoT scenario where several types of sensing applications can be delivered by the satellites. 
In particular, we consider four different possible types of sensing tasks (each related to a different sensed data). 
Each constellation is composed of a given number of satellites, each owning a single resource (but the results can be easily generalized to the case of CubeSats with multiple payloads). We consider both the case of constellations that are either homogeneous or heterogeneous, with reference to the resource type owned by the satellites belonging to the constellation. 

Table \ref{tab:simparameters} summarizes the main simulation parameters. 

\begin{table}[h]
\caption{Main Simulation Parameters.}
\label{tab:simparameters}
\centering
{
\begin{tabular}{|c|c|}
\hline \hline
\textbf{Parameter} & \textbf{Value} \\ \hline \hline
Number of constellations 	& [5-20] \\ \hline
Number of satellites per constellation & [10-60] \\ \hline
Orbital height [Km] & [500-1000] \\ \hline
Number of resources per satellite & 1 \\ \hline
Number of sensing types & 4 \\ \hline
Number of task requests & [100-500] \\ \hline
Execution time & [50-200] ms \\ \hline
Number of runs per simulation & 500 \\ \hline
Compromise factor & 0.5 \\ \hline \hline
\end{tabular}
}
\end{table}

\begin{figure*}[h!]
\centering
\subfigure[Homogeneous constellations - 20 satellites]{\includegraphics[width=0.4\textwidth]{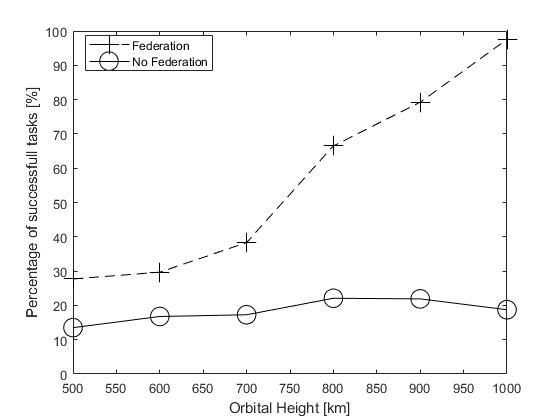}}
\label{fig:varOH20_homo}
\subfigure[Non-homogeneous constellations - 20 satellites]{\includegraphics[width=0.4\textwidth]{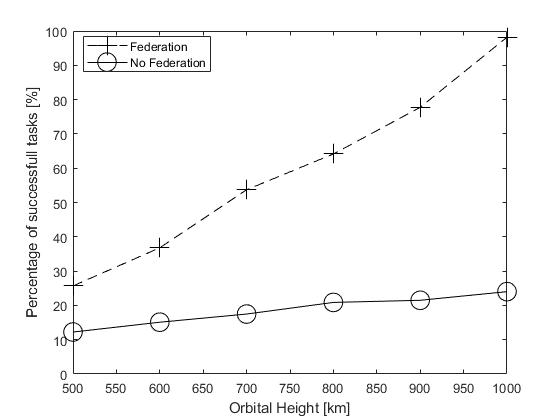}}
\label{fig:varOH20_nhomo}
\subfigure[Homogeneous constellations - 60 satellites]{\includegraphics[width=0.4\textwidth]{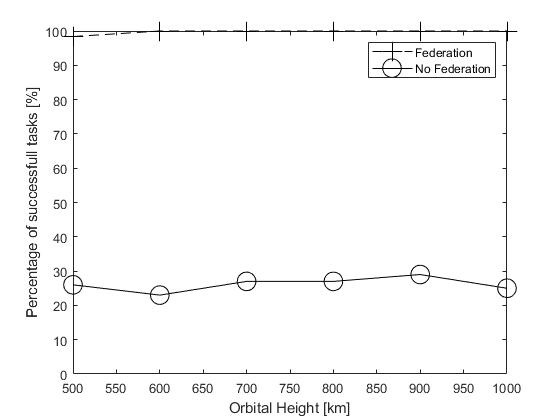}}
\label{fig:varOH60_homo}
\subfigure[Non-homogeneous constellations - 60 satellites]{\includegraphics[width=0.4\textwidth]{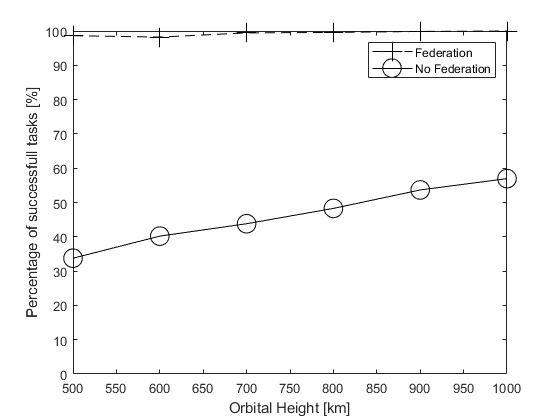}}
\label{fig:varOE60_nhomo}
\caption{Percentage of successful tasks under increasing orbital height in case of homogeneous (a and c) and non-homogeneous (b and d) satellites (100 tasks, 20 constellations).}
\label{fig:varOE}
\end{figure*}

Specifically, the results shown in this paper refer to five scenarios, representative of the typical operational configurations of the platform: 
\begin{itemize}
    \item \textit{Scenario A - Variable orbital height:} the number of required tasks is fixed to 100, the number of virtual constellations is set to 20, and the orbital height varies from 500 to 1000 Km. We consider both the case of 60 and 20 satellites per constellation and of homogeneous and non-homogeneous constellations;
    \item \textit{Scenario B - Variable number of constellations:} the number of satellites per constellation is fixed to 40, the number of required tasks is set to 200, while the number of constellations varies from 5 to 20. We consider the case of homogeneous constellations;
    \item \textit{Scenario C - Variable number of satellites per constellation:} the number of constellations is set to 20, the number of required tasks to 100, and the orbital height is fixed to 500 Km, while the number of satellites per constellation varies from 10 to 60. We consider the case of homogeneous constellations;
    \item \textit{Scenario D - Variable number of required tasks:}  the number of satellites per constellation is set to 40 and the number of constellations to 5, while the number of required tasks ranges from 50 to 300. We consider the case of homogeneous constellations;
    \item \textit{Scenario E - Differentiated required tasks:}  the number of satellites per constellation is set to 40, the number of constellations to 20, satellites orbit at 500 Km above the Earth, four different sensing types are considered, the number of required tasks is 200 while the load of per-type sensing tasks varies. We consider the case of homogeneous constellations.
\end{itemize}

In our analysis, the performance of the proposed federation policy is always compared to the one of a legacy approach that does not allow cooperation among virtual constellations. In particular, we assume that the GS randomly assigns the required tasks to virtual constellations. In the plots, the curve labeled as \textit{No Federation} refers to the scenario in which each virtual constellation executes the assigned tasks by exploiting only its own resources, while the performance of the proposed approach is represented by the curve referred to as \textit{Federation}.

\subsubsection{Scenario A - Variable orbital height}

\begin{figure}[]
\centering
\includegraphics[width=0.4\textwidth]{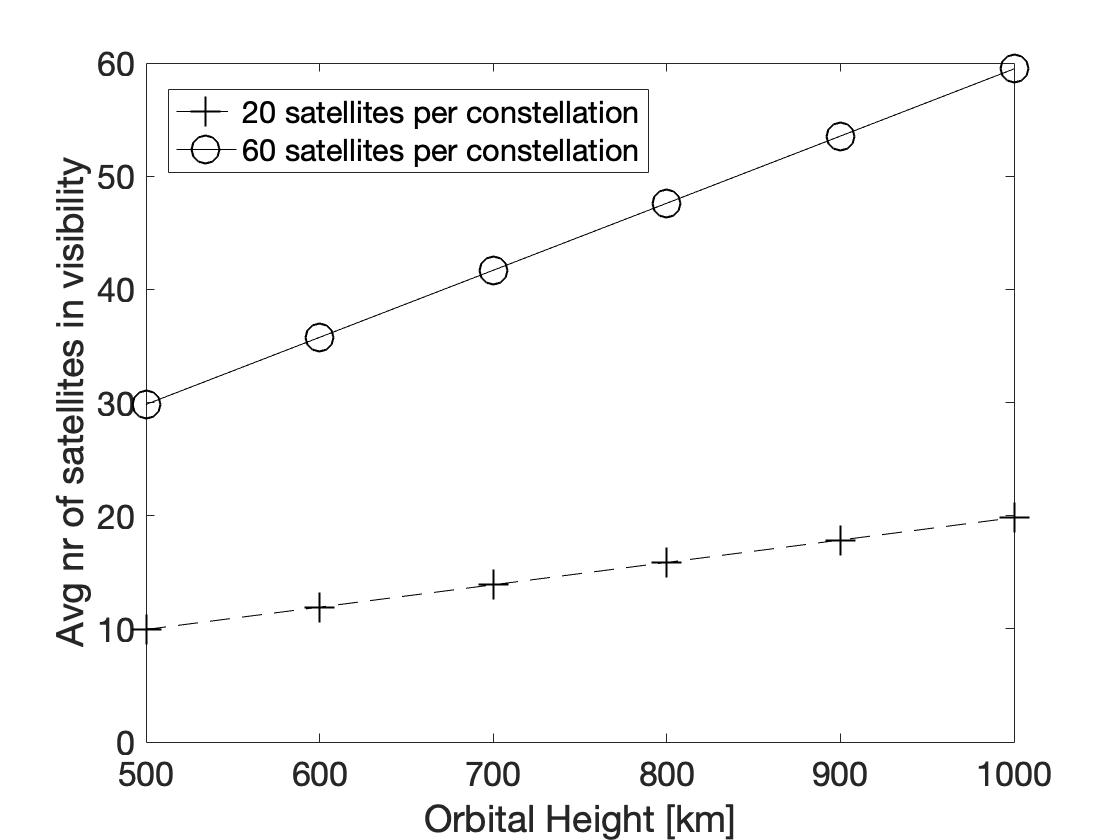}
\caption{Average number of satellites in visibility under increasing orbital height.}
\label{fig:sat_vis}
\end{figure}

Fig. \ref{fig:varOE} shows the percentage of successful tasks when increasing the orbital height in case of homogeneous ((a) and (c)) and non-homogeneous ((b) and (d)) constellations, when considering 20 ((a) and (b)) and 60 ((b) and (c)) satellites per constellation. Plots highlight that the proposed federation policy allows to achieve a substantial performance increase that leads to an amount of successfully executed tasks up to five-time higher than the benchmark approach.
By observing Fig. \ref{fig:varOE}, we can also appreciate that there is a difference in the achieved performance when varying the type of constellation. In particular, while a similar performance is achieved in the cases of \textit{homogeneous} and \textit{non-homogeneous} constellations with 20 satellites per constellation (see (a) w.r.t. (b)), we observe that with the \textit{No Federation} policy the number of successful tasks is lower in the case of 60 satellites per constellation (see (c) w.r.t. (d)). The reason of this behaviour is due to the number of satellites in visibility to the GS. Fig. \ref{fig:sat_vis} reports the average number of satellites in visibility in both the cases of 20 and 60 satellites per constellation analyzed in Fig. \ref{fig:varOE}. We can see that, in the case of 20 satellites per constellation, the GS is in view of no more than one satellite per constellation even when considering the highest orbital height. For this reason, no difference can be appreciated between \textit{homogeneous} and \textit{non-homogeneous} constellations. Instead, in the case of 60 satellites per constellation, always more than one satellite per constellation is in visibility with the GS. Hence, in the case of \textit{non-homogeneous} constellations, the \textit{No Federation} policy is able to complete a higher number of tasks with respect to the \textit{homogeneous} case, since there is a higher probability that the satellites in view own the resources required to complete the assigned tasks. 

As a consequence, the implementation of our proposed \textit{Federation} policy can provide greater benefit in  scenarios characterized by  \textit{homogeneous} constellations and when several satellites, belonging to the same constellation, are in view of the GS.

\begin{figure}[h!]
\centering
\includegraphics[width=0.4\textwidth]{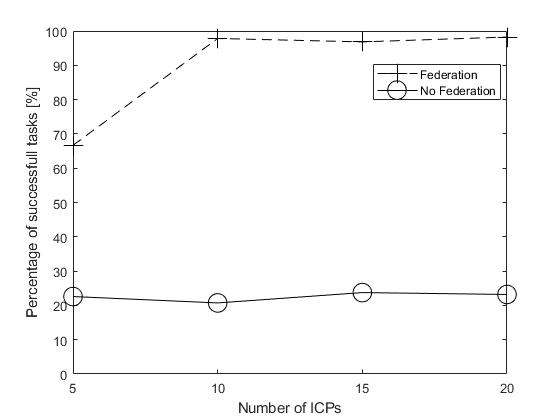}
\caption{Percentage of successful tasks under increasing number of constellations (40 sat per constellation, homogeneous satellites, 200 tasks, 500 Km orbital height).}
\label{fig:varICP}
\end{figure}

\subsubsection{Scenario B - Variable number of constellations}

In Fig. \ref{fig:varICP}, we represent the number of successfully executed tasks under an increasing number of constellations orbiting at 500 Km above the Earth. The number of satellites per constellation is fixed to 40 while the task load (i.e., number of tasks) is set to 200. Plots highlight that the benefits that the \textit{No Federation} policy can derive for an increased number of constellations is limited. Indeed, less than 30\% of the required tasks are executed. On the contrary, the \textit{Federation} policy is able to execute almost all required tasks starting from 10 constellations orbiting around the Earth.

\begin{figure}[h!]
\centering
\includegraphics[width=0.4\textwidth]{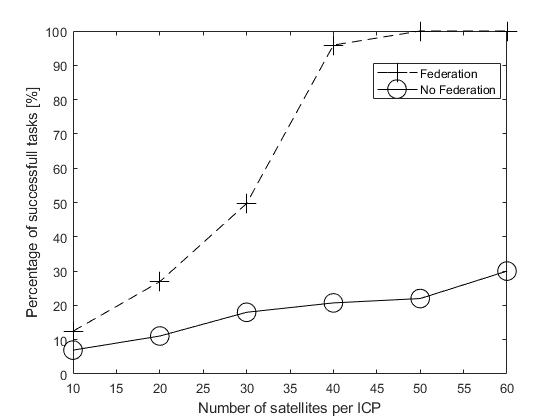}
\caption{Percentage of successful tasks under increasing number of satellites per constellation (20 constellations, homogeneous satellites, 100 tasks, 500 Km orbital height).}
\label{fig:varSat}
\end{figure}

\subsubsection{Scenario C - Variable number of satellites per constellation}

Fig. \ref{fig:varSat} depicts the results achieved in the scenario composed by 20 constellations orbiting at 500 Km above the Earth. Each constellation is composed by a variable number of homogeneous CubeSats (ranging from 10 to 60), which are required to solve 100 tasks. We can appreciate that, in the analyzed conditions, our proposed \textit{Federation} policy is able to solve all required tasks starting from constellations composed by 50 satellites, while only up to 30\% tasks can be solved using the standard approach.

\begin{figure}[h!]
\centering
\includegraphics[width=0.4\textwidth]{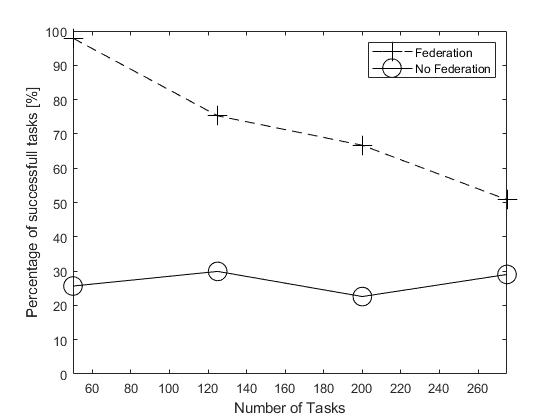}
\caption{Percentage of successful tasks under increasing number of tasks (5 constellations, 40 satellites, homogeneous satellites, 500 Km orbital height).}
\label{fig:varTasks}
\end{figure}

\begin{figure*}[t!]
\centering
\subfigure[No Federation]{\includegraphics[width=0.4\textwidth]{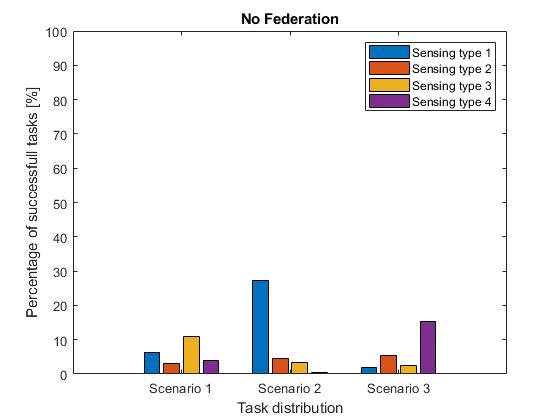}}
\label{fig:varOH20_homo}
\subfigure[Federation]{\includegraphics[width=0.4\textwidth]{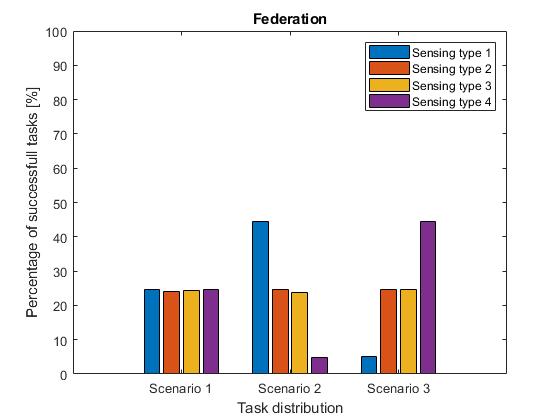}}
\label{fig:varOH20_nhomo}
\caption{Percentage of successful tasks per task type in the three analyzed scenarios (200 required tasks, 40 satellites per constellation, 20 constellations, 500 Km orbital height).}
\label{fig:varExTime}
\end{figure*}

\subsubsection{Scenario D - Variable number of required tasks}

Fig. \ref{fig:varTasks} depicts the results achieved when the task load increases up to 300 required tasks. Differently from the performance of the \textit{No Federation} policy that is not able to reach values higher than 30\%, our proposed \textit{Federation} strategy is able to execute all tasks until the task load is about 50. Under a higher task load, the proposed solution is not able to execute all requested tasks. However, it offers a substantial performance improvement with respect to the \textit{No Federation} policy by achieving at least 20\% gain in terms of completed tasks. A similar behaviour has been observed also under different simulation conditions.
The almost constant trend of the \textit{No Federation} policy is due to the inability to satisfy requests of tasks that require resources not owned by the constellation. On the contrary, the \textit{Federation} policy satisfies all requests until the task load is not too high.

\subsubsection{Scenario E - Differentiated required tasks}

In the last evaluation campaign, we evaluate the performance of the \textit{Federation} policy by considering different types of required sensing tasks (that differ in the time required for their execution), while varying the load of the task types.
In particular, we consider the following four sensing tasks:
\begin{itemize}
    \item Sensing type 1 - 50 ms execution time;
    \item Sensing type 2 - 100 ms execution time;
    \item Sensing type 3 - 150 ms execution time;
    \item Sensing type 4 - 200 ms execution time.
\end{itemize}

Considered sensing types could represent, for example, the request of sensing data (sensing type 1) rather than a high-resolution image (sensing type 4).
We consider that a total of 200 tasks is required and analyze the following three scenarios:
\begin{itemize}
    \item Scenario 1: equal distribution of sensing tasks (25\% for each considered type);
    \item Scenario 2: high load of high-demanding sensing tasks (45\% of sensing type 4, 5\% of sensing type 1, and 25\% of sensing type 2 and 3);
    \item Scenario 3: low load of high-demanding sensing tasks (5\% of sensing type 4, 45\% of sensing type 1, and 25\% of sensing type 2 and 3).
\end{itemize}

Fig. \ref{fig:varExTime} shows that, differently from the \textit{No Federation} policy where the performance remains limited to up to 40\% in the most favorable scenario (i.e., Scenario 2 where most of the tasks to be executed is less demanding in terms of 
execution time), our \textit{Federation} strategy is able to always achieve performance close to 100\%. \newline

In conclusion, we can state that the proposed \textit{Federation} strategy has the potential to bring significant benefits to the 5G IoT-Satellite system consisting of ViCubeSat constellations, at the level of both the system itself and of the single constellation. This is possible owing to the interesting feature of federating, which allows constellations to share their resources virtualized at the network Edges with the common utility of executing as many tasks as possible to considerably enhance the overall system performance.

It can be noticed that our federating solution offers improvements with respect to the legacy approach in terms of percentage of tasks successfully executed even with a few satellites orbiting the Earth at a low altitude, a few constellations (i.e., both homogeneous and non-homogeneous constellations), and a high task load.

\section{Conclusion} \label{Conclusions}

This paper investigated the design of a terrestrial segment of a network for CubeSats that uses typical 5G technologies (SDN, NFV, device virtualization) in order to ``smartly'' integrate this class of satellites into a global IoT network. To this end, a virtual object model to represent CubeSats, based on the OMA Standard, and a platform based on the use of container technology in Multi-Access Edge Computing environments have been also designed. Some scenarios for integrated use of different CubeSats in multi-tenant platforms and effective orchestration of services have been presented and discussed. Finally, results were presented to give an idea on the feasibility of the proposed approach. 


\section*{Acknowledgment}

The authors would like to thank the student Domenico Laganà for his support in the simulation analysis.

This work has been supported by the RUDN University Strategic Academic Leadership Program and PM3 Project (Italian Ministry of University and Research PON Grant nr. ARS01\_01181).


\ifCLASSOPTIONcaptionsoff
  \newpage
\fi




\begin{thebibliography}{1}

\bibitem{cianca_15}
M. De Sanctis, E. Cianca, G. Araniti, I. Bisio, and R. Prasad, ``Satellite Communications Supporting Internet of Remote Things,'' IEEE Internet of Things Journal, 3(1), Feb. 2016.

\bibitem{chong}
S. Laverty, W. C. Chong, J. Osborne, M. Mitry, and V. Lewis, ``The Kepler Satellite System,'' Handbook of Small Satellites, Springer, 2020.

\bibitem{Leo-Sat}
Z. Qu, G. Zhang, H. Cao, and J. Xie, ``LEO satellite constellation for internet of things,'' IEEE Access, vol. 5, pp. 18391-18401, Aug. 2017.

\bibitem{3gpp1} 
3GPP TR 38.811, ``Study on New Radio (NR) to support non terrestrial networks,” Release 15, Sept. 2019.

\bibitem{Woellert}
K. Woellert, P. Ehrenfreund, A. J. Ricco, and H. Hertzfeld, ``CubeSats: cost-effective science and technology platforms for emerging and developing nations,'' Adv. Space Res., 47(4), pp. 663–684, Feb. 2011.

\bibitem{Almon}
V. Almonacid and F. Laurent, ``Extending the coverage of the internet of things with low-cost nanosatellite networks,'' Acta Astronautica, vol. 138, pp. 95-101, May 2017.

\bibitem{puig}
J. Puig-Suari, C. Turner, and W. Ahlgren, ``Development of the standard CubeSat deployer and a CubeSat class picosatellite,'' IEEE Aerospace Conference, Mar. 10-17, 2001. 

\bibitem{sdn-nfv}
V. Nguyen, A. Brunstrom, K. Grinnemo and J. Taheri, ``SDN/NFV-Based Mobile Packet Core Network Architectures: A Survey," in IEEE Communications Surveys \& Tutorials, vol. 19, no. 3, pp. 1567-1602, 2017.

\bibitem{tarik18} 
P. Porambage, et al. ``Survey on multi-access edge computing for internet of things realization,'' IEEE Communications Surveys \& Tutorials, 20(4), pp. 2961-2991, 2018.

\bibitem{mifaas1}
I. Farris, L. Militano, M. Nitti, L. Atzori, and A. Iera, ``MIFaaS: A Mobile-IoT-Federation-as-a-Service Model for dynamic cooperation of IoT Cloud Providers,'' Future Generation Computer Systems, vol. 70, pp. 126–137, 2017.

\bibitem{4} 
N. Fernando, S. W. Loke, and W. Rahayu, ``Mobile Cloud Computing:A Survey,” Future Generation Computer Systems, 29(1), 2013.

\bibitem{Aky1}
Ian F. Akyildiz, and Ahan Kak, ``The Internet of Space Things/CubeSats: A ubiquitous cyber-physical system for the connected world,'' Computer Networks, vol. 150, pp. 134-149, Feb. 2019.

\bibitem{Aky}
I. F. Akyildiz, J. M. Jornet, and S. Nie, ``A new CubeSat design with reconfigurable multi-band radios for dynamic spectrum satellite communication networks,'' Ad Hoc Networks, vol. 86, pp. 166-178, Apr. 2019.

\bibitem{cubesat}
M. Long, et al., ``A cubesat derived design for a unique academic research mission in earthquake signature detection,'' AIAA Small Satellite Conference, 2002.

\bibitem{survey}
Nasir Saeed, Ahmed Elzanaty, Heba Almorad, Hayssam Dahrouj, Tareq Y. Al-Naffouri, and Mohamed-Slim Alouini, ``CubeSat Communications: Recent Advances and Future Challenges,'' IEEE Communications Surveys \& Tutorials, 22(3), third quarter 2020.

\bibitem{134} S. Xu, X. Wang, and M. Huang, ``Software-defined next-generation satellite networks: Architecture, challenges, and solutions”, IEEE Access, vol. 6, pp. 4027–4041, 2018.

\bibitem {Ara-HAP}
G. Araniti, A. Iera, and A. Molinaro. ``The role of HAPs in supporting multimedia broadcast and multicast services in terrestrial-satellite integrated systems,'' Wireless Personal Communications, 32(3), pp. 195-213, 2005.

\bibitem{3gpp2} 
3GPP, TS 36.440, ``General aspects and principles for interfaces supporting Multimedia Broadcast Multicast Service (MBMS) within E-UTRAN,'' Rel. 14, 2017. 

\bibitem{survey_ntn}
F. Rinaldi, H.-L. M\"a\"att\"anen, J. Torsner, S. Pizzi, S. Andreev, A. Iera, Y. Koucheryavy, and G. Araniti, ``Non-Terrestrial Networks in 5G \& Beyond: A Survey,'' IEEE Access, vol. 8, 2020. 

\bibitem{5G_Sat}
A. Guidotti, A. Vanelli-Coralli, M. Conti, S. Andrenacci, S. Chatzinotas, N. Maturo, B. Evans, A. Awoseyila, A. Ugolini, T. Foggi, L. Gaudio, N. Alagha, and S. Cioni, ``Architectures and Key Technical Challenges for 5G Systems Incorporating Satellites,'' IEEE Transactions on Vehicular Technology, 2019.

\bibitem{zimmer}
O. Kodheli, N. Maturo, S. Chatzinotas, S. Andrenacci, and F. Zimmer, ``On the Random Access Procedure of NB-IoT Non-Terrestrial Networks,'' 10th Advanced Satellite Multimedia Systems Conference (ASMS) and 16th Signal Processing for Space Communications Workshop (SPSC), Oct. 20-21, 2020.

\bibitem{comp_24}
E. Zeydan and Y. Turk, ``On the Impact of Satellite Communications over Mobile Networks: An Experimental Analysis,'' IEEE Transactions on Vehicular Technology, 2019.

\bibitem{alagha19} 
N. Alagha, ``Satellite air interface evolutions in the 5G and IoT era,'' ACM SIGMETRICS Performance Evaluation Review, 46(3), pp. 93-95, 2019.

\bibitem{SAGIN}
J. Liu, Y. Shi, Z. M. Fadlullah, and N. Kato, ``Space-Air-Ground Integrated Network: A Survey,'' IEEE Communications Surveys \& Tutorials, 2018.

\bibitem{SAT5G}
K. Liolis, A. Geurtz, R. Sperber, D. Schulz, S. Watts, G. Poziopoulou, B. Evans, N. Wang, O. Vidal, B. T. Jou, M. Fitch, S. D. Sendra, P. S. Khodashenas, and N. Chuberre, ``Use cases and scenarios of 5G integrated satellite-‐terrestrial networks for enhanced mobile broadband: The SaT5G approach,'' International Journal of Satellite Communications and Networking, 2018.

\bibitem{NTN_NR_Ericsson}
X. Lin, B. Hofstr\"om, E. Wang, G. Masini, H.-L. M\"a\"att\"anen, H. Ryd\'en, J. Sedin, M. Stattin, O. Liberg, S. Euler, S. Muruganathan, S. Eriksson G., and T. Khan, ``5G New Radio Evolution Meets Satellite Communications: Opportunities, Challenges, and Solutions'', Computer Science, 2019.

\bibitem{vol19}  
F. Völk, et al. ``Satellite integration into 5G: Accent on first over-the-air tests of an edge node concept with integrated satellite backhaul,'' Future Internet, 11(9), 2019.

\bibitem{wang18} 
N. Wang, et al. ``Satellite support for enhanced mobile broadband content delivery in 5G,'' IEEE International Symposium on Broadband Multimedia Systems and Broadcasting (BMSB), 2018.

\bibitem{Frank}
R. Bassoli, F. Granelli, C. Sacchi, S. Bonafini, and F. H. Fitzek, ``CubeSat-Based 5G Cloud Radio Access Networks: A Novel Paradigm for On-Demand Anytime/Anywhere Connectivity,'' IEEE Vehicular Technology Magazine, 15(2), pp. 39-47, 2020.

\bibitem{Taleb1}
T. Taleb, A. Ksentini, and R. Jantti, ``Anything as a Service for 5G Mobile Systems,'' IEEE Network, 30(6), pp. 84-91, 2016.

\bibitem{Foukas}
X. Foukas, et al., ``Network slicing in 5G: Survey and challenges,'' IEEE Communications Magazine, 55(5), pp. 94-100, May 2017.

\bibitem{Bon12} 
F. Bonomi, R. Milito, J. Zhu, and S. Addepalli, ``Fog computing and its role in the internet of things,'' MCC workshop on Mobile cloud computing, ACM, pp. 13–16, 2012.

\bibitem{Mour17} 
C. Mouradian, D. Naboulsi, S. Yangui, R. H. Glitho, M. J. Morrow, and P. A. Polakos, ``A Comprehensive Survey on Fog Computing: State-of-theart and Research Challenges,'' IEEE Communications Surveys \& Tutorials, 2017.

\bibitem{dong}
X. Zhu, C. Jiang, L. Kuang, M. Dong, and Z. Zhao, ``Capacity Analysis of Multi-layer Satellite Networks,'' The 16th International Wireless Communications \& Mobile Computing Conference (IWCMC 2020), Virtual Conference, Jun. 15-19, 2020.

\bibitem{decola}
T. de Cola and I. Bisio, ``QoS Optimisation of eMBB Services in Converged 5G-Satellite Networks,'' IEEE Transactions on Vehicular Technology, 69(10), pp. 12098-12110, 2020.

\bibitem{Bil18} 
K. Bilal, et al. ``Potentials, trends, and prospects in edge technologies: Fog, cloudlet, mobile edge, and micro data centers,'' Computer Networks, vol. 13, pp. 94-120, 2018.

\bibitem{Yu18} 
W. Yu, F. Liang, X. He, W. G. Hatcher, C. Lu, J. Lin, and X. Yang, ``A Survey on the Edge Computing for the Internet of Things,'' IEEE Access, 6, pp. 6900-6919, 2018.

\bibitem{Chi16} 
M. Chiang, and T. Zhang, ``Fog and IoT: An overview of research opportunities,'' IEEE Internet of Things Journal, 3(6), pp. 854-864, 2016. 

\bibitem{Nit16} 
M. Nitti, V. Pilloni, G. Colistra, and L. Atzori, ``The virtual object as a major element of the internet of things: a survey,'' IEEE Communications Surveys \& Tutorials, 18(2), pp. 1228-1240, 2016.

\bibitem{ETSI} 
ETSI, ``Network Function Virtualization: Architectural Framework,'' 2013. 

\bibitem{noi-bruschi} 
L. Atzori, et al. ``SDN \& NFV contribution to IoT objects virtualization,'' Computer Networks, vol. 149, pp. 200-212, 2019.

\bibitem{Omn18} 
N. Omnes, M. Bouillon, G. Fromentoux, and O. Le Grand, ``A programmable and virtualized network \& IT infrastructure for the internet of things: How can NFV \& SDN help for facing the upcoming challenges,'' IEEE International Conference on Intelligence in Next Generation Networks (ICIN), 2015.

\bibitem{Ojo16} 
M. Ojo, D. Adami, and S. Giordano, ``A SDN-IoT architecture with NFV implementation,'' IEEE Globecom Workshops (GC Wkshps), 2016.

\bibitem{li17} 
J. Li, E. Altman, and C. Touati, ``A general SDN-based IoT framework with NVF implementation,'' ZTE communications, 13(3), pp. 42-45, 2015. 

\bibitem{135} 
A. Kak, E. Guven, U. E. Ergin, and I. F. Akyildiz, “Performance evaluation of SDN-based Internet of space things”, in Proc. IEEE Globecom Workshops (GC Wkshps), pp. 1–6, Dec. 2018.

\bibitem{136} 
T. Li, H. Zhou, H. Luo, and S. Yu, “SERvICE: A software defined framework for integrated space-terrestrial satellite communication,'' IEEE Transactions on Mobile Computing, 17(3), pp. 703–716, Mar. 2018.

\bibitem{Ahmed}
T. Ahmed, E. Dubois, J. B. Dupé, R. Ferrús, P. Gélard, and N. Kuhn, ``Software-defined satellite cloud RAN,'' International Journal of Satellite Communications and Networking, 36(1), pp. 108-133, 2018.

\bibitem{Bertaux}
L. Bertaux, et al., ``Software defined networking and virtualization for broadband satellite networks,'' IEEE Comm. Mag., 53(3), pp. 54-60, 2015.

\bibitem{Li}
T. Li, H. Zhou, H. Luo, Q. Xu, and Y. Ye, ``Using SDN and NFV to implement satellite communication networks,'' IEEE International Conference on Networking and Network Applications, Jul. 23-25, 2016.

\bibitem{Ferrus}
R. Ferrus, et al., ``On the virtualization and dynamic orchestration of satellite communication services,'' IEEE Vehicular Technology Conference (VTC-Fall), Sept. 18-21, 2016.

\bibitem{guo}
X. Zhu, C. Jiang, L. Kuang, Z. Zhao, and S. Guo, ``Two-Layer Game Based Resource Allocation in Cloud Based Integrated Terrestrial-Satellite Networks,'' IEEE Transactions on Cognitive Communications and Networking, 6(2), Jun. 2020.

\bibitem{noi-globecom}
G. Araniti, G. Genovese, A. Iera, A. Molinaro, and S. Pizzi, ``Virtualizing Nanosatellites in SDN/NFV enabled Ground Segments to Enhance Service Orchestration,'' IEEE Globecom 2019.

\bibitem{new1}
J. Long, C. Li, L. Zhu, J. Liu, and Y. Kang, ``Satellite Cloud Architecture Based on Resource Virtualization Technology,'' International Conference on Intelligent Computation Technology and Automation (ICICTA), 2018.

\bibitem{oma-registry}
www.openmobilealliance.org/wp/OMNA/LwM2M/LwM2MRegistry.html

\bibitem{Prado}
J. Prado, ``OMA lighweight M2M resource model,'' IAB IoT Semantic Interoperability Workshop 2016.

\bibitem{geoforall}
www.geoforall.it/k4cdw

\bibitem{openvolcano}
R. Bruschi, et al., ``Openvolcano: An open-source software platform for fog computing,'' International Teletraffic Congress, Sept. 12-16, 2016.

\bibitem{OSCORE}
G. Selander, J. Mattsson, and F. Palombini, ``Object Security for Constrained RESTful Environments (OSCORE),'' draft-ietf-core-object-security-15, August 2018.

\bibitem{gametheory}
W. Saad, Z. Han, M. Debbah, A. Hjørungnes, and T. Basar, ``Coalition game theory for communication networks: A tutorial,'' IEEE Signal Processing Magazine, Special issue on Game Theory in Signal Processing and Communications, Sept. 2009.

\bibitem{stk}
http://www.agi.com/products/engineering-tools

\bibitem{LB}
O. Popescu, ``Power Budgets for CubeSat Radios to Support Ground Communications and Inter-Satellite Links,'' IEEE Access, Jun. 2017.

\end{thebibliography}
%

\end{document}